\documentclass{PoS}
\usepackage{graphicx}
\title{New outbursts of the black hole candidate H1743-322/IGR J17464-3213 observed by INTEGRAL}

\ShortTitle{H1743-322 new outburst}

\author{Fiamma Capitanio\\
        IASF-roma INAF, Rome, Italy, Via Fosso del Cavaliere 100, I-00133 Rome, Italy\\ 
	School of Physics and Astronomy, University of Southampton, Highfield Southampton, SO17 1BJ, UK\\
        E-mail: \email{fiamma.capitanio@iasf-roma.inaf.it}}
\author{A. Bazzano, P. Ubertini\\
        IASF-roma INAF, Rome, Italy, Via Fosso del Cavaliere 100, I-00133 Rome, Italy}
\author{A.J. Bird\\
        School of Physics and Astronomy, University of Southampton, Highfield Southampton, SO17 1BJ, UK
	}
\abstract{On March 2003, INTEGRAL/IBIS detected an outburst from a new source, IGR J17464-3213, that turned out to be a HEAO-1 transient, namely H1743-322. %The spectral and temporal evolution of the source were observed by INTEGRAL in different periods. %Also RXTE observed the source for the first time on 2003 March 29 during a PCA Galactic bulge scan.
Its flux decayed below the RXTE PCA sensitivity limit in November 2003. On July 3, 2004 the source was again detected by RXTE/PCA reaching an intensity of $\sim$70 mCrab in the 2-10 keV band. Another outburst was observed in August 2005. A multiwavelength observation campaign was performed, during the three outbursts, by INTEGRAL, RXTE and CHANDRA satellites and the radio telescopes VLA and ATCA. INTEGRAL data show a good coverage during the July 2004 and August 2005 outbursts.  We show here the spectral and temporal variability study performed on the data collected by INTEGRAL. }
\FullConference{VI Microquasar Workshop: Microquasar and Beyond\\
		September 18-22 2006\\
		Societ\`a del Casino, Como, Italy}
\begin{document}
\section{Introduction}
\begin{figure}[h!]
\centering
 \includegraphics[angle=90, scale=0.3]{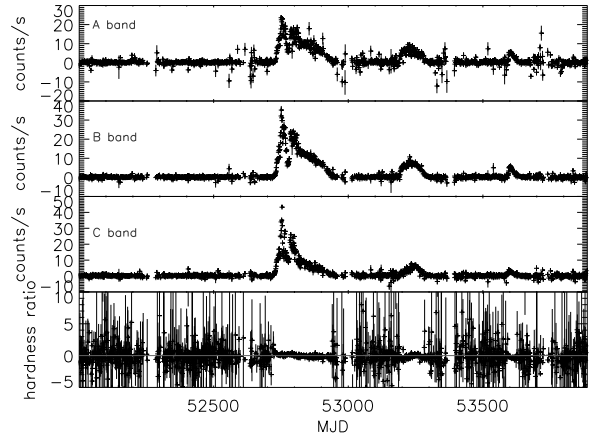}
 \caption{PCA/RXTE light curve in A (1.5-3 keV), B (3-5 keV), C (5-12 keV) energy bands of H1743-322 during its three outburst (from March 2003 to August) and the hardness ratio defined as: $HR=({flux_{C}-flux_{A})/(flux_{A}+flux_{C}})$}
       \label{rxte_abc}
      %    \hspace{1cm}
\end{figure}

\begin{figure}[h!]%[50cm]
 \centering
 \includegraphics[angle=0, scale=0.25]{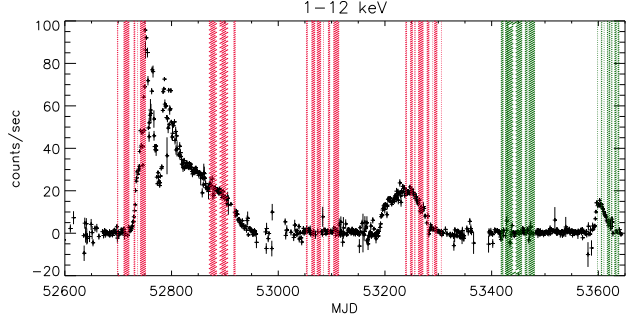}
 \caption{PCA/RXTE 1-12 keV light curve of the H1743-322 temporal behaviour from March 2003 to August. The dotted rectangles are the INTEGRAL observation periods of the source (red:public data, green :proprietary data)}
   \label{integ-rxte}
\end{figure}

{\bf IGR J17464-3213} is associated with H1743-322, a bright black hole candidate (BHC) observed by HEAO1 in 1977. After the start of the outburst in 2003, the source remained bright in soft X rays (E $<$ 15 keV) for $\sim 8$ months and was regularly detected with the JEM-X monitor on board INTEGRAL. At high energy, the IBIS telescope detected the source only on September 9 (52891 MJD). The source, after the first and brightest outburst, showed another two peculiar outbursts. The results of the data analysis of the first and brightest outburst were published in 2005 \cite{Capit2}\cite{Joinet}. A radio-band observation campaign was also performed. A radio flare was seen by NRAO during the rising part of the first outburst (MJD=52728) and two jets were detected between the first and the second outburst (MJD= 52955--53049) \cite{Corb}. For the second outburst, the analysed data show that the source had a short transition to hard state and then back to soft state with a peculiar temporal and spectral behaviour as the RXTE/ASM hardness ratio shows in bottom panel of Figure~\ref{rxte_abc} (Capitanio et al. in preparation). Interestingly, the time period between the peak of an outburst and the subsequent one is equal and it has a value of about 400 days with a decreasing flux. This behaviour suggests a fourth outburst could occur in middle of September 2006. Figure \ref{integ-rxte} shows an indication of the INTEGRAL observation coverage compared with the monitoring of the sources that has been made by  RXTE/ASM from March 2003 to August 2005. 

%
    %2
\section{Time evolution}%\label{time}
\begin{figure}[h!]
\centering
 \includegraphics[angle=90, scale=0.3]{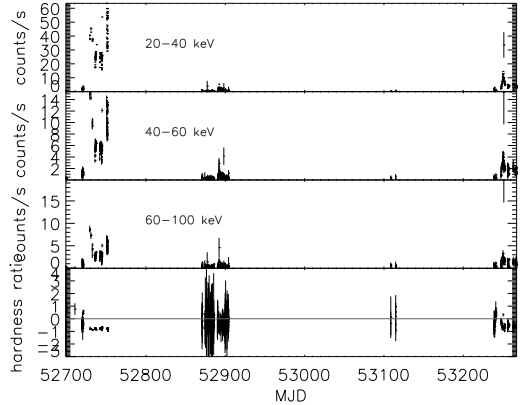}
\caption{IBIS light curves of H1743-322 (20--40 keV, 40-60 keV, 60--100 keV) and the hardness ratio defined as $HR=({flux_{60-100}-flux_{20-40})/(flux_{60-100}+flux_{20-40}})$}
\label{ibistot1}
\end{figure}

\begin{figure}
\centering
\includegraphics[angle=0, scale=0.5]{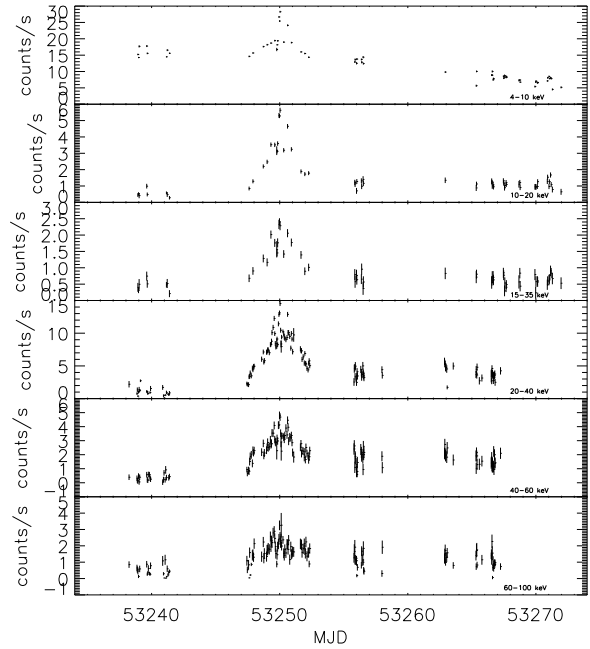}
 \caption{IBIS and JEM--X light curves binned to 2000 seconds of the July 2004 outburst}
 \label{ibis_peak}
\end{figure}

Figure~\ref{rxte_abc} shows the RXTE/ASM light curves in three energy bands (A-B-C) and the hardness ratio. The three peaks of the source outbursts are clearly evident. The first peak was observed by INTEGRAL quite continuously \cite{Capit2}, while the second peak was observed only in its declining phase. Figure ~\ref{ibistot1} shows the IBIS light curves and the corresponding hardness ratio, of all the INTEGRAL public observations of the source. The source spectral evolution of the first outburst followed the typical behaviour of a transient black hole: it was firstly detected in hard state, after passed thorough soft and very high soft state and, at the end of the outburst, it was back to hard state. Also the time evolution of the {\it flux vs photon index} diagram followed the "circular" or hysteresis-like behaviour expected for a transient BHC \cite{Capit2}. Figure~\ref{ibis_peak} shows the JEMX-IBIS light curves in six energy bands of the declining phase of the second outburst. It is noticeable the presence of a small peak around MJD=53250, not evident in the RXTE one-day averaged light curves.
\begin{figure}[h!]
 \centering
  \includegraphics[angle=0, scale=0.5]{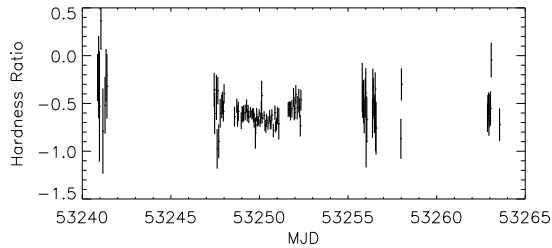}
 \caption{Hardness ratio curve around the second peak. The HR is defined as in Figure 3}%~\ref{ibistot}.}  
    \label{HR_ib}
\end{figure}
%\vspace{-0.5cm}
\section{Spectral evolution}
\begin{figure}[h!]
\centering
\includegraphics[angle=-90, scale=0.3]{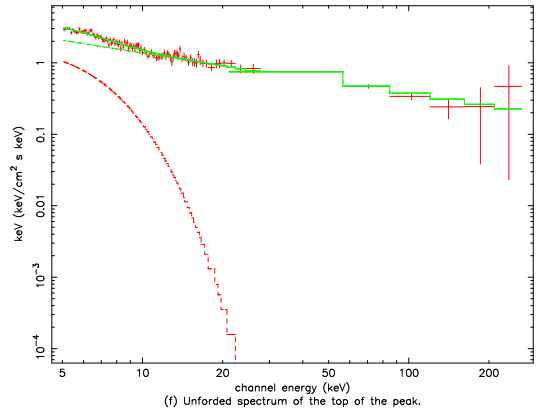}
\caption{Spectrum of the top of the peak shown in Figure 4}
\label{spettro_top}
 \end{figure}

\begin{figure}[h!]
\centering
\includegraphics[angle=0, scale=0.3]{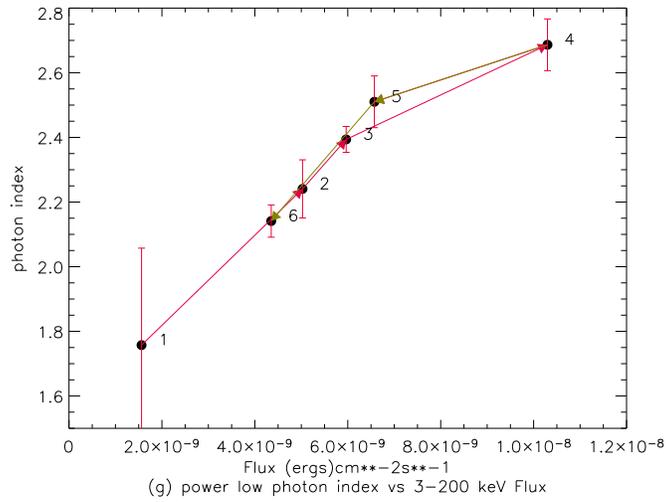}
\caption{Photon index vs flux of the different spectra during the outburst}
\label{pho_vs_flux}
\end{figure}

The spectral analysis presented here is focused on the secondary peak detected by INTEGRAL shown in Figure~\ref{ibis_peak}. The source spectrum varies from the start to the end of the peak. The spectra are all well fitted with a disc black body component plus a power law without a cutoff. The spectrum shown in Figure~\ref{spettro_top} represents the  peak of the light curves in Figure~\ref{ibis_peak}. The spectral behaviour indicates, as the hardness ratio (Figure~\ref{HR_ib}), a softening of the source with the presence of a moderate black body component ($T_{in}$=1.2 keV) and a hard tail without any cutoff up to 200 keV. The diagram in Figure~\ref{pho_vs_flux} shows the photon index variation of the power law component versus flux (3-200 keV). The diagram shows that the source became softer and brighter and then went back to a harder state creating a sort of loop in the diagram.

\section{Conclusions}
As seen in Figure~\ref{integ-rxte} the INTEGRAL observations cover only the final part of the outburst. Looking at the RXTE hardness ratio (Figure~\ref{rxte_abc} bottom panel) it is noticeable that, contrary to the other two outbursts (the first and the third ones), the second outburst of the source starts with an hardening and then comes back to the soft state. The secondary short 4 days peak, detected by INTEGRAL, is not visible in the RXTE one day averaged light curve. The source, during this short period, has a sharp transition to soft state. The best fit for the high energy data is a power law without any detectable cutoff even near the top of the peak, where the source is bright enough to permit the extraction of the IBIS spectrum up to 200 keV. This suggests the presence of non thermal processes at work. 

\end{document}